\documentclass[a4paper,11pt]{article}
\pdfoutput=1 

\usepackage{jcappub} 

\usepackage[T1]{fontenc} 
\usepackage{graphicx,caption,amsmath, amssymb} 
\usepackage{threeparttable}
\usepackage{slashed}
\usepackage{footnote}
\usepackage[utf8]{inputenc}

\def\beq{\begin{equation}}
\def\eeq{\end{equation}}
\def\barr{\begin{array}}
\def\earr{\end{array}}
\def\dis{\displaystyle}
\def\tev{\, {\rm TeV}}
\def\gev{\, {\rm GeV}}
\def\lapp{\mathrel{\rlap{\raise.5ex\hbox{$<$}}
                    {\lower.5ex\hbox{$\sim$}}}}
\def\gapp{\mathrel{\rlap{\raise.5ex\hbox{$>$}}
                    {\lower.5ex\hbox{$\sim$}}}}

\title{\boldmath Universal Extra Dimensions and the Graviton Portal to Dark Matter}

\author[a,b]{Mathew Thomas Arun,}
\author[c]{Debajyoti Choudhury,}
\author[c,1]{Divya Sachdeva\note{Corresponding author.}}

\affiliation[a]{Department of Physics, Mar Thoma College, Thiruvalla 689 103, Kerala, India}
\affiliation[b]{Center for High Energy Physics, Indian Institute of Science, Bangalore-560 012, India}
\affiliation[c]{Department of Physics and Astrophysics,University of Delhi, Delhi 110 007, India}

\emailAdd{thomas.mathewarun@gmail.com}
\emailAdd{debajyoti.choudhury@gmail.com}
\emailAdd{divyasachdeva951@gmail.com}

\abstract{The Universal Extra Dimension (UED) paradigm is
    particularly attractive as it not only includes a natural candidate for the Dark Matter particle
    , but also addresses several issues related to particle
    physics. Non-observations at the Large Hadron Collider, though,
    has brought the paradigm into severe tension.  However, a particular 
  5-dimensional UED model emerges from a six dimensional space-time
  with nested warping. The $AdS_6$ bulk protects both the Higgs mass
  as well as the UED scale without invoking unnatural parameter
  values.  The graviton excitations in the sixth direction open up new
  (co-)annihilation channels for the Dark Matter particle, thereby
  allowing for phenomenological consistency, otherwise denied to the
  minimal UED scenario. The model leads to unique signatures in
    both satellite-based experiments as well as the
    LHC.\\ \textbf{Keywords: Dark Matter, UED, Graviton, Warped Compactification}}

\begin{document}
\maketitle
\flushbottom

\section{Introduction}
The presence of dark matter (DM) in the universe and its dominance
over luminous matter is
well-established~\cite{Komatsu:2010fb,Ade:2015xua}.  What is not
apparent, though, is its nature, a consequence of the lack of any
evidence (direct or indirect) barring the astrophysical/cosmological
context.  Furthermore, the absence of any DM candidate within the
Standard Model (SM) requires us to propose scenarios going beyond.
The primary requirement is to have at least one component that is
stable over cosmological times scales. Very often, this stability is
guaranteed by a $Z_2$ symmetry, with the DM (and, possibly, its
heavier cousins) being odd under it and all SM particles being even.

It would be particularly appealing if any such theory can also address
at least some of the several other outstanding issues in the SM. Most
notable of the latter are the hierarchy problem, the mechanism of
generating the baryon asymmetry in the universe, the flavor problem
(including the large hierarchy in the fermion masses), the unification
of forces, etc.. Two particular classes that address most (if not all)
of these issues are provided by supersymmetry and/or extra-dimensional
scenarios. Here, we concentrate on the latter alternative.

The
RS-model~\cite{Antoniadis:1990ew,Gogberashvili:1998vx,Randall:1999ee}
provided a particularly elegant resolution of the hierarchy, between
the quantum-gravity scale $M_5$ and the electroweak scale $v$.
through $v = \tilde v \, \exp(- \pi \, k_5 \, r_c)$, where $\tilde v =
{\cal O}(M_5)$ and $k_5$ denotes the bulk curvature, and $r_c$ is the
stabilized value of the modulus.  Some of the aforementioned issues
(as also proton decay and FCNC) can be addressed too if the SM gauge
fields and fermions are not restricted to the TeV brane (as in the
original RS model), but are promoted to 5-dimensional
ones. Constraints from the electroweak precision tests as well as
other low-energy observables necessitates the introduction of
additional symmetries and fields, though.  Existence of a viable
DM-candidate requires even further additions. More damaging, though,
is the non-observation of any Kaluza-Klein (KK) excitation of the
graviton at the LHC, namely the constraint $m_1 \gapp 4\tev$ (at
95\%C.L.)  ~\cite{Khachatryan:2016yec,ATLAS-CONF-2016-018} On the
other hand, the model implies $m_n = x_n \, k_5 \, \exp(- \pi \, k_5
\, r_c)$, where $x_n$ are the roots of the Bessel function of order
one. With the applicability of semi-classical arguments~\cite{Davoudiasl:1999jd} (upon which
the model hinges) as well as arguments relating the D3 brane tension
to the string scale (and, hence, to $M_5$ through Yang-Mills gauge
couplings) restricting $k_5/M_5 \lapp 0.15$, one would, thus, expect
the first KK-mode to be, at best, a few times heavier than the Higgs,
and the non-observation implies at least a little hierarchy.

The Universal Extra Dimension (UED)
scenario~\cite{Appelquist:2000nn,Appelquist:2000nn2}, on the other
hand, envisages a flat compactified fifth dimension of radius $R$ and
a $S^1/Z_2$ orbifolding (thereby ensuring that zero mode fermions are chiral in nature) 
with all the SM fields allowed to propagate in the bulk.
Although KK-number is broken, in the absence of any brane-localized
terms in the Lagrangian, a $Z_2$-subgroup (``KK-parity'') is retained.
Given by $(-1)^n$ where $n$ is the KK-level, this renders the lightest
KK-excitation absolutely stable. With the mass of the $n^{\rm th}$
mode of a species (with a five-dimensional mass $m_0$) being given by
\begin{equation}
m_n^2 = m_0^2 + n^2/R^2 \ ,
\label{UED_spectrum}
\end{equation}
clearly the excitations of a given level are nearly degenerate for
 $R^{-1} \gapp 1\tev$. However, quantum corrections (due to both the
 bulk fields as well as the orbifolding) provide additional splitting,
 resulting in $B^{(1)}$ (the first excitation of the hypercharge
 boson) being the lightest, and, hence, the DM
 candidate~\cite{darkued1,darkued12,darkued13,darkued14}. With all
 interactions being determined by the SM gauge action, the pair
 production of the KK-excitations at colliders is unsuppressed
 (barring kinematics), leading to the possibility of striking
 signatures\footnote{Note that, in such models, there is no warping
 and, hence, the gravitons have only suppressed ($M_{\rm Pl}^{-1}$)
 coupling to the rest, rendering them irrelevant to low-energy
 processes.}. The non-observation of such states at the LHC, whether
 it be monojet searches at $\sqrt{s} =
 8\tev$~\cite{CMS8_monojet,ATLAS8_monojet} or multijets at $\sqrt{s} =
 8\tev$~\cite{TheATLAScollaboration:2015sgw,CMS13_multijets}, has been
 used to impose constraints of $R^{-1} \gapp
 1150 \gev$~\cite{Choudhury:2016tff}. Subsequent
 simulations~\cite{Deutschmann:2017bth,Beuria:2017jez} suggest that
 the bound can be strengthened to $R^{-1} \gapp 1400 \gev$. On the
 other hand, agreement with the observed DM energy
 density~\cite{Komatsu:2010fb,Ade:2015xua} suggests that $R^{-1} \lapp
 1400 \gev$~\cite{Belanger:2010yx}, an upper bound that cannot be
 relaxed as the interactions are well-specified and the
 mass-differences constrained. A further refinement of the search
 strategies, including a count of (soft) particle
 multiplicities~\cite{c_n_s} would impose even stronger constraints.
 Thus, we are approaching an era of
 tension and the model stands to be comprehensively ruled
 out\footnote{It should be noted, however, that the non-minimal version can alter the spectrum, 
 thereby possibly evading the problem, by invoking tuned brane localized and/or higher dimensional terms in the lagrangian.
 The existence of such terms is anticipated, 
 as the UED is not a UV-complete theory but only an effective field theory with a cutoff not much larger than $R_y^{-1}$. 
 A symmetric character of such terms(necessary for stability of the DM) is to assured by {\em imposing} a KK-parity.}. 
 A further, and unrelated, problem with
 the UED paradigm is that there is no mechanism to stabilize the
 modulus $R$, thereby leaving it free to change with time, assuming
 any value.

\section{Model}
To solve both these problems in a unified manner, we consider a
six-dimensional space-time with successive (nested) warpings along the
two compactified dimensions, which are individually $Z_2$-orbifolded
with 4-branes sitting at each of the edges. 
The uncompactified directions ($x^\mu)$  support
four-dimensional Lorentz symmetry. With
the compact directions represented by the angular coordinates $x_{4,5}
\in [0,\pi]$ and the corresponding moduli by $R_y$ and $r_z$,   the line
element is, thus, given by~\cite{Choudhury:2006nj}
\begin{equation}
\label{metric}
ds^2_6= b^2(x_5)[a^2(x_4)\eta_{\mu\nu}dx^{\mu}dx^{\nu}+R_y^2dx_4^2]+r_z^2dx_5^2 \ ,
\end{equation} 
where $\eta_{\mu \nu}$ is the flat metric on the 
four-dimensional slice of spacetime.

Denoting the fundamental scale in six dimensions by $M_6$ and the negative (six dimensional) bulk cosmological constant by 
$\Lambda_6$, the total bulk-brane action is, thus,
\begin{equation}
\barr{rcl}
{\cal{S}}&=& \dis {\cal{S}}_6+{\cal{S}}_5 \\[1.5ex]
{\cal{S}}_6&=& \dis \int d^4x \, dx_4 \, dx_5 \sqrt{-g_6} \, 
   (M_{6}^4R_6-\Lambda_6)\\[1.5ex]
{\cal{S}}_5&=& \dis \int d^4x \, dx_4 \, dx_5 \sqrt{-g_5}\, 
      [V_1(x_5) \, \delta(x_4)+V_2(x_5) \, \delta(x_4-\pi)]\\
&+& \dis \int d^4x \, dx_4 \, dx_5\sqrt{-\tilde g_5} \, 
     [V_3(x_4) \, \delta(x_5)+V_4(x_4) \, \delta(x_5-\pi)] \ . 
\earr
\end{equation}
The five-dimensional metrics in ${\cal S}_5$ are those induced on the
appropriate 4-branes which accord a rectangular box shape to the
space. 

  The action above, augmented by the choice of the line
  element as in eqn.(\ref{metric}) leads to straightforward Einstein
  equations.  However, before listing them, it should be pointed out
  that the solutions as presented in the original paper,
  viz. Ref.\cite{Choudhury:2006nj}, were not the most general
  ones. Rather, they had assumed specific values for the induced
  five-dimensional cosmological constants on the 4-branes (and,
  similarly, for the four-dimensional cosmological constants on the
  3-branes).  Relaxing this allows for more generic solutions,
  including the existence of bent branes. For example, if an induced
  four-dimensional cosmological constant $\Omega$ were to be admitted,
  the four dimensional components of the Einstein equations  would
read~\cite{Arun:2016csq,Choudhury:2006nj}
\[
a^2 \left[\frac{3}{R_y^2}\left(\frac{a''}{a}+\frac{a'^2}{a^2}\right)
             +\frac{2}{r_z^2}\left( 3 \dot b^2 + 2 b \ddot b
	           +\frac{\Lambda_6 \, r_z^2}{2 M_6^4}b^2\right)\right] 
             =\frac{\Omega}{r_z^2} \ ,
\]
where primes(dots) denote derivatives with respect to $x_4$ ($x_5$).
The aforementioned induced
cosmological constant (on a five dimensional
hypersurface along constant ($x_5$) makes an appearance in the form of
 a constant of separation $ \widetilde{\Omega} $, 
leading to
\beq
\label{bequation}
3\dot b^2 + 2 b \ddot b +\frac{\Lambda_6 \, r_z^2}{2 M_6^4}b^2 = \widetilde{\Omega}
\eeq
and
\beq
\label{aequation}
\frac{3}{R_y^2}\left(\frac{a''}{a}+\frac{a'^2}{a^2}\right)
     +\frac{2}{r_z^2}\widetilde{\Omega} = \frac{\Omega}{r_z^2 \, a^2}\, .
\eeq
Only if the 
induced five-dimensional cosmological 
constant\footnote{Note that the metric on a constant-$x_5$
slice is given by
\[
ds^2 = e^{-2c |x_4|}\eta_{\mu \nu}dx^{\mu}dx^{\nu} + R_y^2 dx_4^2 \ .
\]
For such a $AdS_5$ space, we know that $c = \frac{R_y}{2}\sqrt{-\Lambda_5/12 M_5^3}$,
where $\Lambda_5$ is the (induced) cosmological constant on the 
5 dimensional manifold and $M$ the fundamental energy scale on it. 
Using the second of eqns.(\ref{generic_solz}) and eqn.(\ref{generic_sol}),
we have $\widetilde\Omega = r_z^2\Lambda_5 /4 M_5^3$. 
In other words, although $\widetilde{\Omega}$ makes an appearance 
as a separation constant in the equation of motion, it is inherently 
related to the induced cosmological constant
on any $x_5=$ constant hyper-surface.}
 $\widetilde \Omega$ vanishes 
identically, does the nonlinear differential equation for $b(x_5)$ simplify to yield a special class of solutions~\cite{Arun:2016csq}
that are untenable for $\widetilde \Omega \neq 0$. However, rather than limit 
ourselves to this fine-tuned case, we consider the  generic case. It is 
interesting to note, though, that the two solutions are very similar
differing, at most, by  50\% (almost independent of the value of $\widetilde \Omega$)~\cite{Arun:2016csq}.
Assuming\footnote{For $\widetilde \Omega > 0$, one has, instead, 
$b(x_5) =  b_1 \, \sinh(k |x_5| + b_2)$. Now, $b_2 = 0$ 
would no longer be allowed unless one is willing to admit a vanishing metric, 
albeit only for $x_5 = 0$. The rest of the phenomenology would be quite 
analogous to the present case.} $\widetilde \Omega < 0$, the first equation has the solution 
\beq
\barr{rclcl}
\label{generic_solz}
b(x_5) & = & \dis b_1 \cosh(k |x_5| + b_2) \\[2ex] b_1 & = & \dis
\sqrt{\frac{-\widetilde \Omega}{3 \, k^2}} & = & \dis {\rm sech}(k \pi
+ b_2)\\[2ex] k & = & \dis r_z \sqrt{\frac{-\Lambda_6}{10 \, M_6^4}} &
\equiv & \dis r_z \, M_6 \, \epsilon \ , \earr \eeq where the second
equality in the second line is nothing but the normalization $b(x_5 =
\pi) = 1$. The dimensionless constant $\epsilon$ is a measure of the
bulk cosmological constant $\Lambda_6$ in terms of the fundamental
scale $M_6$. Clearly, if $\epsilon$ is too large, the bulk would be
highly warped and a semiclassical treatment (as is being attempted
here) would be invalid.  Typically, it has been argued that $\epsilon
\lapp 0.15$, by arguments relating the brane tension to the scale of
some underlying string theory (or even to
$M_6$)~\cite{Davoudiasl:1999jd}.  Although somewhat larger values can
be admitted, this would entail the applicability of the semiclassical
approximation growing progressively worse. On the other hand, the
aforementioned limit automatically ensures that the curvature in the
$x_4$-direction is sufficiently small.
 
The generic solution to eqn.(\ref{aequation}) for a
nonzero $\Omega$, is also given in terms of hyperbolic functions and
is quite complicated.  The subsequent algebra is rendered extremely
complex and does not offer easy insights. On the other hand, the
observed cosmological constant in our world is infinitesimally small
and we must live close to $\Omega = 0$.  Hence, we shall assume that
$\Omega = 0$. While this may be termed a fine-tuning, it is, at
worst, exactly the same as that in the RS model. Indeed, $\Omega =
0$ is not a special solution, and a similar criticism could be made
against any finite value for $\Omega$. On the other hand, $\Omega =
0$ could, in principle, have resulted from some as yet unspecified
symmetry~\cite{Alexander:2001ic}. In this limit, the solution can
be expressed as

\beq
\label{generic_sol}
a(x_4) = e^{-c|x_4|}  \ , \qquad  
    c \equiv b_1 \, k \, \frac{R_y}{r_z} \ ,
\eeq
where we have normalized by imposing $a(x_4 = 0) = 1$.
The graviton has a tower of
towers. The modes in the $x_4$-direction are given in terms of Bessel functions, while those in the $x_5$-direction are
given in terms of associated Legendre functions~\cite{Arun:2014dga}.  This change,
along with the fact of the hierarchy resolution now being shared
between two warpings, results in the mass of the first KK-mode being
significantly higher than that of the corresponding mode in the RS
case.  Furthermore, the large coupling enhancement that allowed for
the RS-gravitons to be extensively produced at the LHC, is now
tempered to a significant degree~\cite{Arun:2014dga}. Consequently,
the production rates are suppressed and the scenario easily survives
the current bounds~\cite{Arun:2014dga,Arun:2015ubr}.

The brane potentials are determined by the junction conditions. 
The ones at $x_5 = 0, \pi$ are simple  and 
are given by
\beq
V_3 = \frac{-8 M_6^4 k}{r_z} \tanh(b_2) \ , \qquad 
V_4 = \frac{8 M_6^4 k}{r_z} \tanh(k \pi + b_2) \ ,
\eeq
whereas the ones at $x_4 = 0, \pi$ have $x_5$--dependent tensions 
\beq
\label{brane_potential}
V_1(x_5) = - V_2(x_5) = 
 \frac{8 M_6^4 c}{R_y \, b(x_5)} 
   = \frac{8 M_6^4 k}{r_z} \, {\rm sech}(k |x_5| + b_2)\ .
\eeq

It should be noted that the Israel junction condition $V_1 = - V_2$ is
the consequence of choosing $\Omega = 0$, or, in other words, a
configuration wherein the four-dimensional cosmological constant
vanishes exactly. Had we chosen to work with $\Omega \neq 0$, this
equality of magnitude would not have held. This is exactly
  analogous to the case of the vanilla RS model.

\begin{figure}[!ht]
\vspace*{-100pt}
\centerline{
\includegraphics[width=12cm,height=14cm]{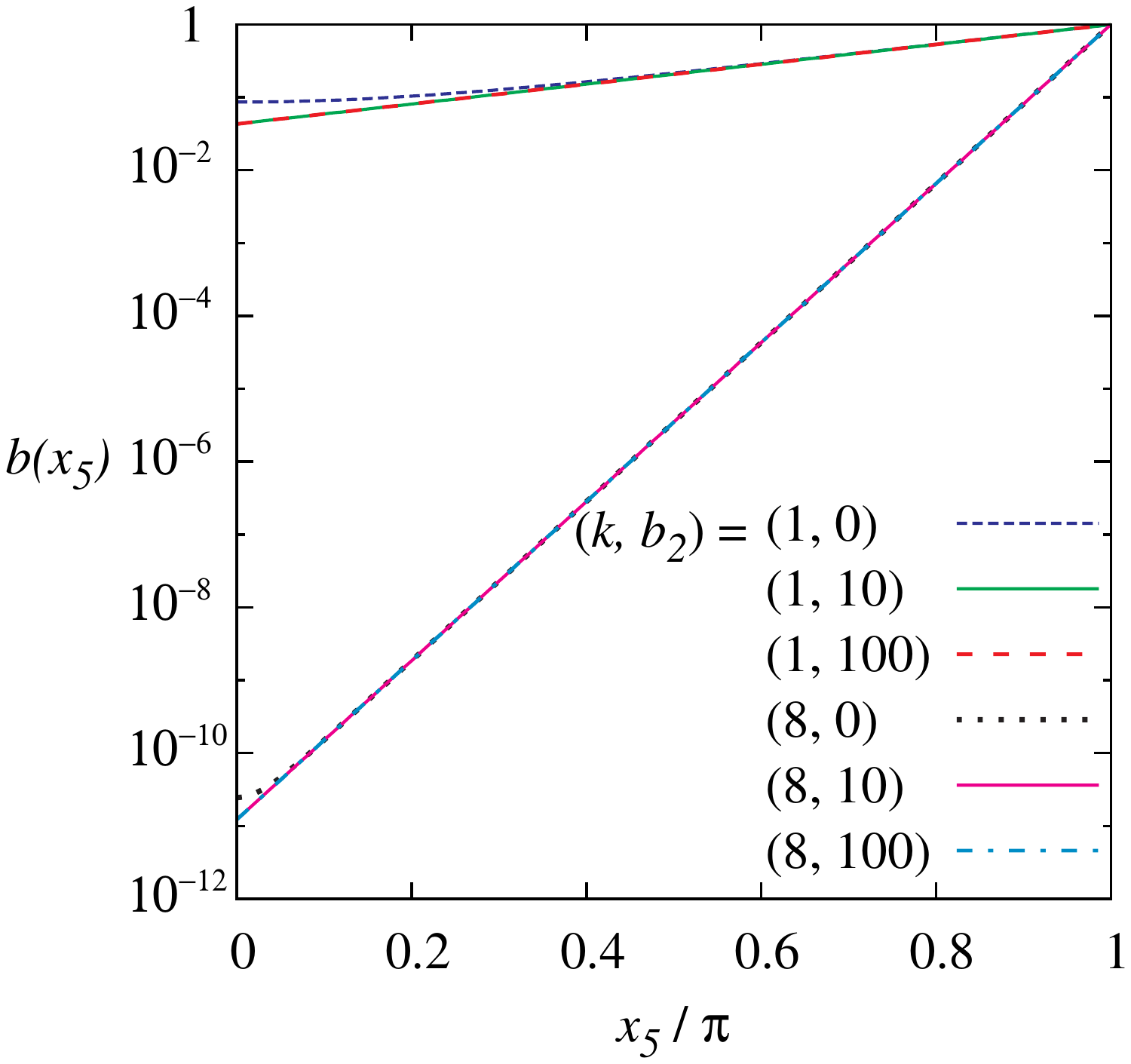}
}
\vspace*{-100pt}
        \caption{\small \em{The warp function $b(x_5)$ for some choices 
of the parameters.}}
        \label{fig:b_func}
\end{figure}
In Fig.\ref{fig:b_func}, we show the metric element $b(x_5)$ for
several choices of the parameters $k$ and $b_2$. It is quite apparent
that, for a given $k$, the warp factor depends very little on
$b_2$. While some difference is still noticeable for small $k$ values,
for slightly larger $k$, the almost marginal difference is
concentrated close to $x_5 = 0$. 

Given this relative insensitivity, it is conceivable that limiting
values of $b_2$ may be used to understand the main features and the
results for intermediate values of $b_2$ may be obtained by
interpolating between such extreme cases. Some of these limits lead to
significant algebraic simplifications, opening the possibility of
closed-form analytic solutions. For example, choosing $b_2 = 0$
recovers the results of Ref.\cite{Choudhury:2006nj} and we
have\footnote{It should be noted here both $c\ll k$ as well as
    $c > k$ are realizable here.}
\beq \barr{rcl} c & = & \dis
\frac{R_y \, k}{r_z} \, {\rm sech}(k \pi) \\[2ex] 
V_1(x_5) &= &\dis -
V_2(x_5) = \frac{8 M_6^4 k}{r_z} {\rm sech}(k |x_5|) \, ,
\\[2ex]
V_3 &= & 0 \\[1ex] V_4 & = & \dis \frac{8 M_6^4 k }{r_z} \, \tanh(k \pi) \, \, .
\earr \eeq 
This corresponds to a bent brane scenario with non-vanishing induced
five-dimensional cosmological constants on the hypersurfaces at $x_5 =
0,\pi$. This could easily be seen by observing that the induced metric
on the $x_5 = 0$ surface, apart from an overall $b(0)$ factor, is
given by
\[
ds_5^2 = e^{- 2 c |x_4|} \eta_{\mu \nu} dx^{\mu}dx^{\nu} + R_y^2 dx_4^2 \, \, ,
\]
or, in other words, the induced geometry is $AdS_5$-like.

\section{Naturally Stabilised UED}

We are particularly interested in the limit opposite to that discussed
at the end of the preceding section, namely, in 
a sufficiently large $b_2$.  For
such values, $b(x_5) \approx (b_1 / 2) \exp(k |x_5| + b_2)$.  For
sufficiently large $b_2$, we have $c \ll k$ (see
eqn.\ref{generic_sol}), no matter how large the ratio $R_y/r_z$ might
be.  With $c \rightarrow 0$ the brane potentials now read
\beq 
\barr{rcl}
V_1 = - V_2 & \approx & 0 
\\[1ex] 
V_3 & \approx & \dis \frac{- 8 M_6^4 k}{r_z}  
\approx - V_4 \ .
\earr
\eeq 
The fact of $V_3 \approx - V_4$ reveals the near vanishing of the
cosmological constant induced on the brane. As for the 
line element,  in this limit, 
\[
ds^2 = e^{2 k (|x_5|-\pi)} \Big( e^{- 2 c |x_4|} \eta_{\mu \nu} dx^{\mu}dx^{\nu} +R_y^2 dx_4^2 \Big) + r_z^2 dx_5^2 
\]
\[
\approx e^{2 k (|x_5|-\pi)} \Big(\eta_{\mu \nu} dx^{\mu}dx^{\nu} +R_y^2 dx_4^2 \Big) + r_z^2 dx_5^2 \ .
\]
This metric is conformally flat and along with the $AdS_6$ bulk, it
resembles a generalization of the Randall-Sundrum geometry in
5-dimensions.  With this simplification, the 4-brane tensions are
rendered pairwise equal and opposite~\cite{Arun:2016csq}, reflecting
the vanishing of the induced cosmological constant on the
branes\footnote{This is not a fine-tuning, as the solutions are
  qualitatively similar, for the generic case as well.  Moreover, the
  4-dimensional cosmological constant is independent of the
  5-dimensional one.}.

  It should be realized, though, that the approximate conformal
  flatness would have followed as long as $c \ll k$. We do not
  strictly need either of $b_2 \to \infty$ or $c \to 0$.  A
  sufficiently large 
  $b_2 \gapp 75$ would lead to $c \lapp 10^{-22}$ (a value that will be shown to be 
    small enough to support a viable DM candidate)
  and qualitatively the same conclusions as a larger $b_2$. 

Choosing a particular value for $b_2$ (this choice also served to
determine $c$) that corresponds to a 
small $\widetilde{\Omega}$
represents a small
 five-dimensional cosmological constant
(equivalently, straight, or unbent, four-branes at the ends of the
world). The opposite limit corresponds to the case wherein the
four-branes suffer the maximum possible bending commensurate with a
semiclassical analysis (or, in other words, a five-dimensional
cosmological constant comparable to the fundamental scale).  The low
energy phenomenology, naturally, would turn out to be quite different
in the two cases. Clearly, any intermediate value of $b_2$ would
correspond to a intermediate value of the five-dimensional
cosmological constant and, similarly, for the low-energy
phenomenology.

We now turn to the issue of stabilization. Contrary to the RS
  case, here we need to stabilize two moduli, namely $r_z$ and
  $R_y$. The former can be trivially stabilized a la
  Goldberger-Wise~\cite{Goldberger:1999uk} through the introduction of
  a bulk scalar with a simple quadratic potential, which, when
  integrated out, provided the requisite effective potential for the
  radion~\cite{Arun:2016csq}. Of course, such a scalar field
  contributes to the energy density of the bulk and the solution for
  the metric would be altered. However, with the space-time now close
  to being conformally flat, $r_z$ can be stabilized (including
  backreaction) through a introduction of a bulk scalar with a quartic
  potential~\cite{Arun:2016csq}, and a closed-form solution for both
  the scalar and the graviton wavefunction obtained.  The resultant
  distortion of the graviton spectrum and the couplings thereof is
  minimal.  With the natural scale on the 4-brane at $x_5 = \pi$ now
  being warped down to the TeV-scale, a scalar localized on this brane
  would be expected to stabilize $R_y^{-1}$ to the same scale. Indeed,
  as Ref.\cite{Arun:2016csq} points out, an analogous exercise can
  also be undertaken towards stabilizing $R_y$ with a scalar now
  introduced on the brane at $x_5 = 0$. While this is technically
  correct, it should be noted that, for values of $c$ that we require
  from the DM perspective, the corresponding potential needs to be a
  very shallow one, thereby calling into question the
  stability. However, this issue can be viewed from a different
  perspective. The very fact that the warping in the $x_5$-direction
  stipulates an energy scale, of ${\cal O}(10 \tev)$ on the brane at
  $x_5 = 0$ implies that any dynamics on this brane must be at
  energies lower than this scale. On the other hand, in the absence of
  any other restraining mechanism, masses would tend to rise to the
  maximum possible scale. In other words, one would expect $R_y^{-1}$
  to lie somewhat below the fundamental energy scale on this brane.

It is also interesting to consider the effect of a rolling
  $R_y$. This can be done by promoting $R_y$ to a dynamical field (in
  the same spirit as a radion is stabilized) and examine its evolution
  as a function of time. While first indications are that this is a
  slow process, we shall desist from a full discussion at this point.

Considering  the SM fields to be
five-dimensional ones, defined on the entire 4-brane at $x_5 = 0$,
leads, now, to a stabilized 5-dimensional UED-like scenario. With the Higgs field being localized on the TeV brane (the 4-brane 
located at $x_5=0$), we have
\begin{equation}
m_{\rm Higgs} = \gamma \, r_z^{-1} \, e^{- k \pi} \ ,
\end{equation}
a relation bearing close resemblance to the 5D one. 
Here, $\gamma/r_z$ is the Higgs mass at the UV brane, 
($r_z^{-1}$ being the natural scale) with $\gamma$ 
encompassing
both the possible small hierarchy between the fundamental parameters 
as well as quantum corrections to the measurable. In delineating parameter 
space, we use $1 > \gamma \gapp 0.7$. Note, here, that the lower limit 
is not sacrosanct, but has been motivated by the aesthetic desire of 
not introducing a large ``little hierarchy''.

\begin{figure}[!ht]
\centerline{
\includegraphics[width=13cm,height=7cm]{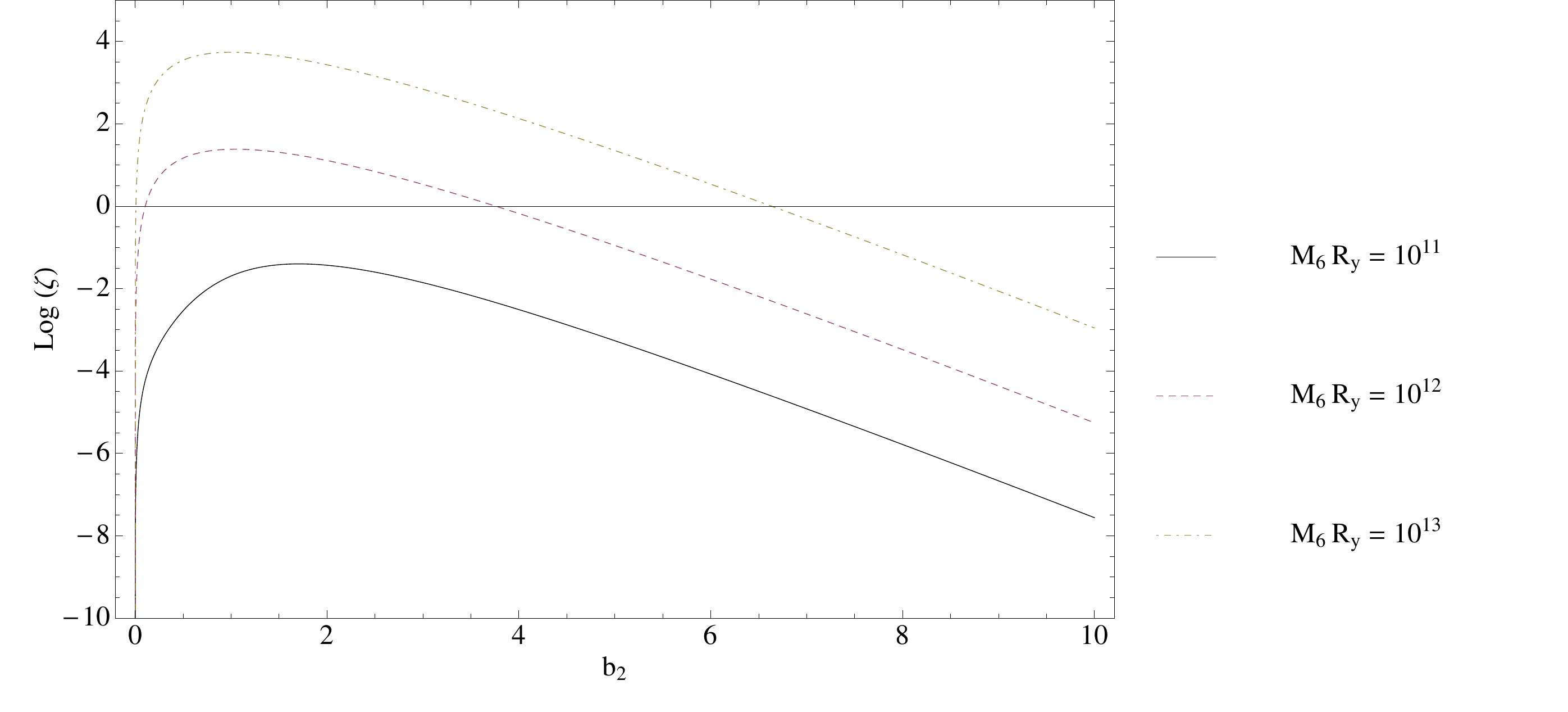}
}
        \caption{\small \em{The Barbieri-Giudice parameter $\zeta$ as a function of $b_2$ for 
        $k =8$, $\epsilon = 0.15$ and three different values of the product 
$M_6 \, R_y$.}}
        \label{BGparam_fig}
\end{figure}

Before we end this section, let us take a look at 
a concrete measure of the sensitivity of the Higgs mass to the 
key parameter, namely $b_2$. The analogue of the well-known 
Barbieri-Guidice naturalness parameter~\cite{Barbieri:1987fn,Ellis:1986yg}
is given by
\begin{equation}
\zeta = \frac{\partial \ln(m_{\rm Higgs}(b_2)/m_0)}{\partial \ln(b_2)} 
   \qquad  m_0 \equiv \gamma \, r_z^{-1} \ .
\label{BGparam}
\end{equation}
Using the relations derived in this section, we present, in
Fig. \ref{BGparam_fig}, this parameter as a function of $b_2$ for a
relevant choice of $(k, \epsilon)$ and three choices of the product $M_6
\, R_y$. It can be seen easily that $\zeta$ is acceptably small for
$b_2$ values of interest ($\gapp 70$). Even though $\zeta$ can become
large for larger values of the product $M_6 \, R_y$, the fine-tuning
becomes unacceptably large only for $b_2 \sim 15$, far away from the
range of interest.

\section{Gravitons and their Interactions}
The (four-dimensional) Planck mass, in such theories is 
a derived quantity rather than a fundamental one and is obtained 
by integrating the action over $x_{4,5}$, thereby yielding
\begin{equation}
M_{\rm Pl}^2 = (M_6^3 R_y \pi / 3 \epsilon) \, \left[1 - \exp(-3 k \pi) \right] \ . 
\end{equation}

With $\epsilon \lapp 0.15$ and $R_y^{-1} \ll M_6$, we clearly 
have $M_6 \ll M_{\rm Pl}$. This is markedly different from the case 
of the original RS model and has considerable phenomenological consequences
\cite{Arun:2014dga,Arun:2016ela}.
Expanding the graviton wavefunction as 
\beq
h_{\mu \nu}(x_\mu,x_4,x_5) = \frac{1}{\sqrt{R_y r_z}}
   \sum_{n,p} h^{(n,p)}_{\mu \nu}(x_\mu) \Psi_{n}(x_4) \chi_p(x_5) \ ,
\eeq
the equation of motion for the $x_5$ component is 
\[
\left[\partial_5(b^5\partial_5) + m_p^2 b^3\right] \chi_p(x_5) = 0 \ ,
\]
with the solution being given in terms of associated Legendre 
functions~\cite{Arun:2014dga}. For large $b_2$ though, 
these simplify~\cite{Arun:2016ela},
in terms of $\tilde{z} \equiv e^{k(\pi-x_5)}$
\beq
\dis \chi_p \approx \dis N_p^{-1} \, \tilde{z}^{5/2}
    \left[J_{5/2}\left(m_p \, r_z \, k^{-1} \tilde{z} \right) 
      + \beta_{p} Y_{5/2}\left(m_p \, r_z \, k^{-1} \tilde{z} \right) \right] \,
\label{grav_wavefn}
\eeq
where  the constants $N_p$ are determined 
from the normalization conditions
\beq
\int_{-\pi}^{\pi} d x_5 \, b^3 \chi_p \chi_{p'} = \delta_{p,p'} \ .
 \label{grav_normaliz}
\eeq
The Neumann boundary conditions ($\chi'_p$ vanishing at both $x_5 = 0,\pi$) 
demand that $\beta_p$ be vanishingly small and 
\[
m_p = k \, e^{-k \pi} \, r_z^{-1} \, x_p
\]
where $x_p$ are the roots of $J_{3/2}(x)$. As for the equation of motion for $\Psi_n(x_4)$, in this regime of small $c$, 
it is almost a flat Laplacian, and we have, for the masses of the 4-dimensional graviton modes $h^{(n,p)}_{\mu \nu}(x_\mu)$,
\[
m_{np}^2 = p_4^2+p_5^2 \approx n^2 \, R_y^{-2} + m_p^2 \ .
\]
and 
\begin{equation}
\Psi_n(x_4) = \frac{1}{2\pi} + \frac{\cos(nx_4)}{\pi} \ .
\label{eq:psi}
\end{equation}
With the $x_4$-direction being nearly flat, $(-1)^n$ is a very good
symmetry for gravitons and the SM-fields alike. With the latter
being expressible in terms of $\sin(n \, x_4)$ and $\cos(n \, x_4)$,
we have, for the graviton interactions,
\begin{equation}
\barr{rcl}
{\cal L}_{\rm int.} &=& \dis \int_{0}^{\pi} \frac{dx_4}{M_6^2}\,
h_{\mu \nu}(x_\mu,x_4,x_5=0) T^{\mu \nu}(x_\mu,x_4)
\earr
\label{eq:coupling}
\end{equation}
where the energy momentum tensor
$T^{\mu\nu}(x_\mu,x_4)$ for gauge field(G) and fermion field(F) is expressible
as
\begin{equation}
\barr{rclcl}
T_G^{\mu \nu}(x_\mu,x_4) &=& \dis \frac{1}{2\pi R_y} T_G^{(0,0)}(x_\mu) 
& + & \dis \sum_{n',m=1} \frac{1}{\pi R_y} T_G^{(n',m)}(x_\mu) \cos(n'x_4) \cos(mx_4)
\\[3ex]
T_F^{\mu \nu}(x_\mu,x_4) &=& \dis \frac{1}{2\pi R_y} T_{Fl}^{(0,0)}(x_\mu) 
& + & \dis \sum_{n',m=1} \frac{1}{\pi R_y} T_{Fl}^{(n',m)}(x_\mu) \cos(n'x_4) \cos(mx_4) 
\\[1ex]&& & + & 
\dis \sum_{n',m=1} \frac{1}{\pi R_y} T_{Fr}^{(n',m)}(x_\mu) \sin(n'x_4) \sin(mx_4) 
\ ,
\earr
\label{eq:emtensor}
\end{equation}
the difference between the left ($l$) and right ($r$) fermion 
modes\footnote{While there exists another term in
  $T_F^{\mu\nu}(x_\mu,x_4)$ mixing left and right fermion modes, it is 
  odd under $x_4 \to -x_4$, and does not contribute on 
    integrating over $x_4$.}
accruing from the different boundary
conditions~\cite{Arun:2016ela}. Using eqns.(\ref{eq:emtensor} \& \ref{eq:psi}) and
integrating the $x_4$ and $x_5$ directions, we get
\begin{equation}
\barr{rcl}
{\cal L}_{\rm int.} & = & \dis \sum_{p,n'=0} 
   C^{(0,p)}_{n'}h^{(0,p)}_{\mu \nu}(x_\mu) \, T^{\mu \nu(n',n')}_i(x_\mu) + \sum_{n,p,n'=1} 
   C^{(n,p)}_{n'}h^{(n,p)}_{\mu \nu}(x_\mu) \, T^{\mu \nu(n',|n\pm n'|)}_i(x_\mu)
\\[2.5ex]
& = & \dis \sum_{n,p,n' = 0}^\infty C^{(n,p)}_{n'}h^{(n,p)}_{\mu \nu}(x_\mu) \, T^{\mu \nu(n',|n\pm n'|)}_i(x_\mu)
\earr
\label{eq:expr1}
\end{equation}
where $i\in G,F$.  For our analysis, the only relevant graviton is the
first excitation in the $x_5$ direction\footnote{While the couplings
  of $h^{(1,0)}_{\mu\nu}$ (indeed, of all $h^{(n>0,0)}_{\mu\nu}$) are
  Planck-suppressed, the other graviton modes like $h^{(n, p>1)}$ are
  too heavy to be of immediate concern. Consequently, the
  corresponding 4-fermion interactions are also highly suppressed, to
  levels well below those relevant for DM (co-)annihilation as well as
  well-measured low-energy processes.} namely
$h^{(0,1)}_{\mu\nu}$. And the only relevant ones if its couplings are
those to the SM fields (namely, $C^{(0,1)}_0$) and their first
KK-modes in the $x_4$ direction (viz. $C^{(0,1)}_1$). These are
actually identical, and we use the compact notation
\begin{equation}
\barr{rcl}
C^{0,1} \equiv C^{(0,1)}_0=C^{(0,1)}_1=\dis 
    \frac{1}{4\pi} \; \frac{\chi_1(x_5=0)}{M_6^2\sqrt{R_y r_z}} \ .
\earr
\label{eq:expr2}
\end{equation}
  Using eqns.(\ref{grav_wavefn} \& \ref{grav_normaliz}), the
  common coupling $C^{0,1}$ can be expressed in terms of the
  parameters of the theory in a straightforward way. What is important
  to note is that $C^{0,1}$ cannot vary immensely as long as the
  extent of the warping is not changed drastically.  This is quite
  similar to the case of the RS scenario where, for the first
  KK-graviton, the (dimensionful) coupling to the SM fields is only
  somewhat smaller than $m_{\rm graviton}^{-1}$. Much the same happens
  here.  There is a small further suppression~\cite{Arun:2014dga}
  owing to two factors. For one, the extent of the hierarchy (between
  the fundamental scale and the IR scale) is somewhat different owing
  to the presence of an extra direction. Moreover the graviton
  wavefunction at $x_5 = 0$ is somewhat modified from the RS case (in
  the case at hand, this transpires in the orders of the Bessel
  functions).


\section{The Relic Density}

Within the mUED, the $B^{(1)}$ turns out to be the lightest of the 
KK-excitations, and owing to the existence of a KK-parity (namely, the $Z_2$ 
symmetry of $x_4 \leftrightarrow -x_4$) is exactly stable. In the present case,
the truth of $B^{(1)}$ being the lightest still continues to hold. 
However, the non-vanishing of $c$ explicitly breaks the KK-parity, and would 
induce decays of the  $B^{(1)}$. The leading role in such decays 
is played by a tree-level KK parity-violating vertex connecting 
the $B^{(1)}$ to a pair of SM fermions, generated on KK-reduction in the 
presence of the slightly warped background.  In the small $c$ limit, 
the corresponding vertex factor is given by 
   \[
 \mathcal{V} \approx - i~g_1 \, Y_f~\gamma_\mu~\frac{\sin(\pi+c~\pi^2/2)}{\pi}
             \approx i \, \frac{g_1 \, c \, \pi}{2} \, Y_f \, \gamma_\mu
   \]
where $g_1$ is the $U(1)_Y$ coupling and $Y_f$ the hypercharge of the 
fermion. The total (fermionic) decay width is, thus, given by 
  \[
    \Gamma_{B^{(1)}\rightarrow {\rm fermions}} \approx
    \frac{\pi \, g_1^2 \, c^2}{48} \; M_{B^{(1)}} \; 
            \sum_f Y_f^2 \, C_f \, \left(1-4\frac{m_f^2}{M_{B_1}^2}\right) 
                 \, \left(1+2\frac{m_f^2}{M_{B_1}^2}\right)
  \]
  where $C_f$ is the number of colors and the sum extends over all
  fermions.  Decays into scalars(operative in the radion higgs mixing) are of no importance.Thus, for the DM
  to be stable over cosmological timescales, one needs $c \lapp
  10^{-22}$. Note that such values of $c$ are engendered by $b_2 \gapp
  75$, a region that is commensurate with desirably low values of the
  Barbieri-Giudice sensitivity measure (see. Fig.\ref{BGparam_fig}).

Owing to their near degeneracy in mass and similarity of couplings,
the generic first KK-excitation does not decouple much earlier than
the $B^{(1)}$. Consequently, they play an important role in
determining the relic density $n_{B^{(1)}}$, not only through co-annihilations,
but also by replenishing $B^{(1)}$ through processes like $e^{(1)+}
e^- \to B^{(1)} + \gamma$. Indeed, for mUED, the latter effect is
strong enough that the inclusion of these excitations, especially the
singlets, forces down $R^{-1}$ by nearly 250
GeV~\cite{Belanger:2010yx,Cornell:2014jza}. The exact magnitudes of
such effect depends on the quantum corrections to the mass splittings,
which, in turn, are determined by the cutoff scale $\Lambda$. Within
the mUED, it has been suggested~\cite{Datta:2012db} that the stability
of the Higgs potential restricts $\Lambda \lapp 5 R^{-1}$, but such
bounds can be evaded. We consider, instead, large
representative values, namely $\Lambda R_y =
  2, 4, 35$, with the first two corresponding to small splittings and
  the last to large ones. It should be realized here that the exact
  value of $\Lambda$ is an imponderable. All that can be said with any
  certainty is that it should be sufficiently larger than the first
  graviton excitation mass and, hence, $2 \, R_y^{-1}$, but not
  orders of magnitude larger\footnote{This can also be 
      viewed thus: the stabilization of $r_z$ implies that the
    two 4-branes are fixed at two locations. Assuming that the 
    hierarchy problem is solved, the UV brane is fixed such that 
    the natural scale on it is $M_6$ whereas the IR brane is 
    to be fixed at the few-TeV scale.  The exact scale, while  
    unknown, can be fixed by choosing a values for $k$ without 
    resorting to extreme fine-tunings. Naturally, all particles 
    localized to this brane should have a mass smaller than
    this scale, whereas particles with larger masses should see the $x_5$ 
    direction. In other words, the delta-function profile for the 
    brane that one has assumed becomes untenable at this energy 
    and would need to be resolved to see the entire structure of this 
    domain wall (brane). In other words, we are faced with an
    five-dimensional effective field theory with a cutoff($\Lambda$) that is 
    somewhat (but not too much) larger than the masses of the first 
    few modes in the $x_5$-direction. This is, then, further compactified 
    down to four dimensions. While the cutoff($\Lambda$) in a UED theory could, 
    conceivably, be high, here it can, at best, equal the
    aforementioned scale.}.

In the stabilized version, the presence of the graviton 
excitations in the $x_5$-direction 
introduces additional diagrams both in the $s$- and $t$-channels (the
latter, particularly, for co-annihilations). However,
given their somewhat smaller couplings, the gravitons (essentially,
only $h^{(0,1)}$) would turn out to really manifest themselves in the
$s$-channel.  For an accurate computation of the relic density (known
to be $\Omega h^2=0.1199\pm
0.0022$~\cite{Komatsu:2010fb,Ade:2015xua}), we implemented our model
(including the quantum-corrected couplings of the KK-2 states to SM
particles~\cite{Belanger:2010yx}) in micrOMEGAs~\cite{Belanger:2006is}
using LanHEP~\cite{Semenov:2014rea}.  As a check, we have compared
against the CalcHEP model file discussed in
Ref.~\cite{Belyaev:2012ai}.

\begin{figure}[!ht]
\vspace*{-40pt}
\centerline{
\includegraphics[width=14cm,height=12cm]{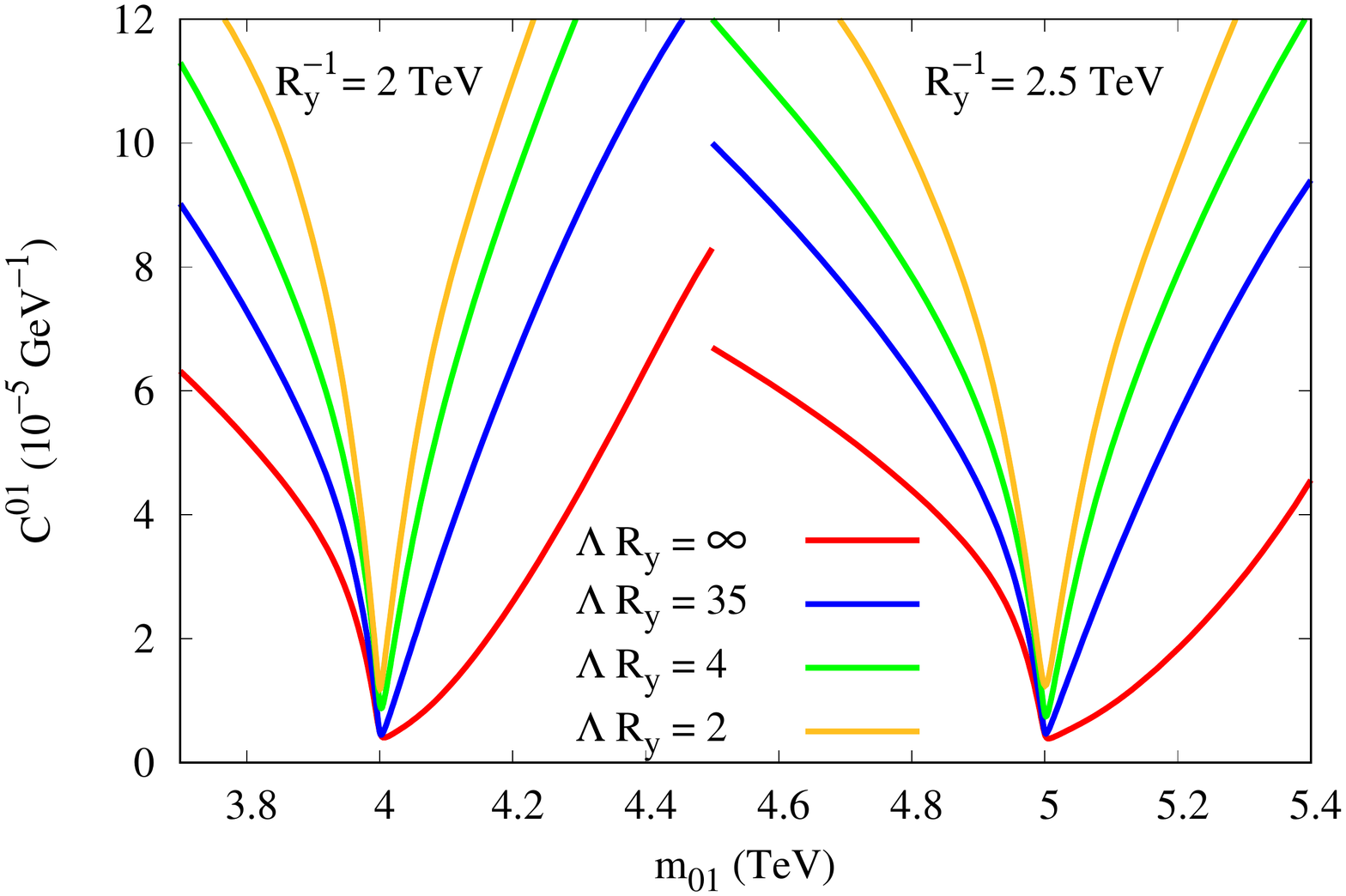}
}
\vspace*{-10pt}
        \caption{\small \em{Relic density contours 
for different values of $R_y$ and $\Lambda$. Parameter points
above the curves are allowed.
}}
        \label{fig:omega}
\end{figure}
With the graviton in play, an efficient annihilation mode opens up,
thereby allowing a much heavier $B^{(1)}$ to be consistent with the
relic density than was allowed within mUED (see Fig.\ref{fig:omega}).
Understandably, this process is most efficient close to the resonance,
which, for the non-relativistic DM particles, would occur when
$m_{01} \approx 2 M_{B^{(1)}} \sim 2 R_y^{-1}$. In
Fig.\ref{fig:omega}, this is attested to by the severe dip in the size
of the required coupling, as is shown by the lowest curve
corresponding to the hypothetical case where all SM excitations other
than the $B^{(1)}$ have been switched off (or $\Lambda R \to
\infty$). The asymmetrical nature of the curve is a result of the
interplay of three factors, the natural width of the $h^{(0,1)}$ which
grows as the cube of its mass, the fact that the $B^{(1)}$ may have a
small but nonzero momentum, and the functional dependence of the
annihilation cross sections.

More realistic cases are represented by the other curves in
Fig.\ref{fig:omega}. While the salient features remain similar, the
increase in the required coupling was expected in view of the
discussions above. The dependence of the increase on $\Lambda$ is but
a reflection of the fact that a smaller $\Lambda$ implies smaller
splittings, and, hence, more states conspiring to increase $n_{B^{(1)}}$. 
Furthermore, as $m_{01}$ grows sufficiently beyond $2M_{B^{(1)}}$, 
the graviton is allowed to decay into the other SM KK-excitations 
as well, thereby ameliorating the asymmetry. For small enough $\Lambda$ 
and large enough $m_{01}$ this effect clearly overcompensates. 

\begin{figure}[!ht]
\vspace*{-40pt}
\centerline{
\includegraphics[width=12cm,height=12cm]{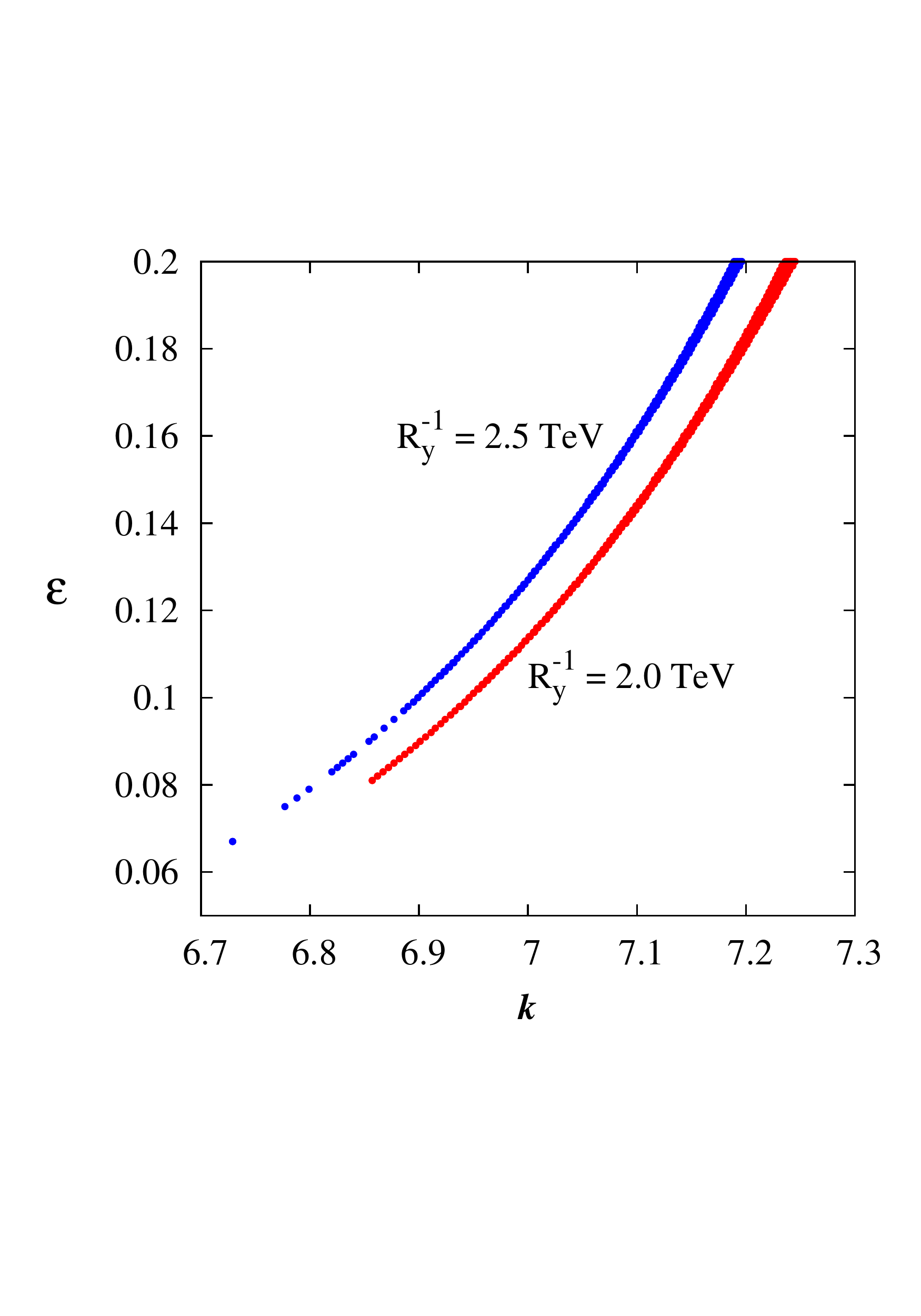}
}
\vspace*{-70pt}
        \caption{\em{The region in the $k$--$\epsilon$ plane allowed by
            the relic density for two values of $R_y$. Difference
            between $\Lambda R = 4 \, (35)$ is invisible on the scale
            of the graph.}}  
\label{fig:param}
\end{figure}

Fig.\ref{fig:param} displays the region in the parameter space that
satisfies the relic density constraint 
as well as reproducing $m_{\rm Higgs} = 126$ GeV without invoking
a substantial little hierarchy ($1 > \gamma \gapp 0.7$). 
For a given compactification radius $R_y$, the
fundamental parameters $k$ and $\epsilon$ are highly correlated,
with the latter adopting moderate values consonant with a
semiclassical treatment.  It is particularly intriguing to note that
the fundamental scale in such models is close to the left-right
symmetry scale, namely $M_6 \sim 10^{14}$ GeV (with $r_z^{-1}$ being
a factor of 20--80 smaller). This is likely to have interesting
implications for model building.

\section{Conclusions and Outlook}

The generalization of the RS paradigm to nested warping, thus, not
only sheds some light on the very scale of the UED
  paradigm (in turn, related to the mass hierarchy), but also
resolves the acute tension between the requirements imposed by
the DM relic density on the one hand, and non-observation
of the UED excitations at the LHC on the other. While the
  first issue is not entirely solved, very good reasons exits why the
  scale of the compactification in the UED direction should be a
  little below that of the inherent energy scale of the UED brane,
  which, in turn, is stabilized through the introduction of bulk
  scalars that provide the requisite effective potentials. The
  solution to the second (and, phenomenologically, more immediate)
  problem is much easier to understand. The immersion of the UED-like
  model in a RS-like warped background (in a higher dimension)
  provides a new mode for DM-DM annihilation. The consequent dilution
  in the relic number density allows for heavier DM and, thus, a
  higher UED scale, thereby allowing one to escape the conflict
  between LHC observations and the bounds from
  WMAP/Planck. Simultaneously, this immersion helps us evade the LHC
  structures against the plain RS scenario. This, essentially, comes
  about as a consequence of two features. For one, the introduction of
  a large quasi-UED scale serves to reduce the fundamental scale from
  $M_{\rm Pl}$ to one that is significantly smaller. Simultaneously,
  going to the sixth dimension results in quantitatively changing the
  wavefunction of the corresponding KK-graviton near the IR brane,
  thereby suppressing its coupling to the brane localized fields. This
  has the direct consequence of suppressing the single production of
  the graviton (which would, then, show up in modes such as dileptons,
  diphotons, dijets etc.). The very same immersion is also the one
  responsible for the protection of the UED scale.

Given these successes, it is contingent upon us to examine the
consistency of this model with other experiments.  As for direct
detection experiments, the smaller couplings (and at least twice as
large a mass) of the gravitons along with the fact that they appear
only in the $t$-channel renders their contribution to be vanishingly
small. The DM-nucleon scattering cross sections, thus, are very
similar to the mUED case, and for the parameter space in
Fig.\ref{fig:param}, explicit calculations (using micrOMEGAs) yield
numbers well below the strongest current limits~\cite{Akerib:2016vxi}.

  The situation is much more complex when it comes to indirect
  detection. The thermal-averaged annihilation cross sections for $B_1
  B_1 \rightarrow b\bar{b}/W^+ W^-/ZZ/gg$ etc. are several orders of
  magnitude below the most restrictive limits from
  AMS-02\cite{Lin:2015taa} and Fermi-LAT\cite{Ackermann:2015zua}. On
  the other hand, the unsuppressed coupling (vis. a vis. other
  species) of the $B_1$ to photons implies a significantly large
  cross-section for a pair of the DM particles annihilating into a
  diphoton pair.  With the latter being quasi-monochromatic, this
  would stand out in the sky. Indeed the corresponding rates are very
  close to the upper limits deduced from H.E.S.S.(consider, for example, Fig.4 of
    Ref.\cite{Abramowski:2013ax}). In particular, if
    parameters are such that a pair of a non relativistic DM particles is 
    very close to the graviton resonance, the rate of annihilation
    could even exceed the limits imposed by H.E.S.S.. For example,
    consider a situation with $R_y^{-1}= 2.5 \tev$ and
    $\Lambda=35$. If we have $m_{01}=5$ TeV ({\em i.e.}, almost on
    resonance), this would result in $\langle\sigma v
    \rangle(B_1B_1\rightarrow\gamma\gamma)=$6.1$\times$
    10$^{-27}$cm$^3/$s, which is larger than what H.E.S.S. allows for.
    On the other hand, if we had $m_{01}=4.7$ TeV instead, then (even
    while accounting for the larger coupling required to satisfy the
    relic density bound), we would have had $ \langle\sigma
    v\rangle(B_1B_1\rightarrow\gamma\gamma)=$3.6$\times$
    10$^{-27}$cm$^3/$s, and, hence, well within the allowed
    region. Much the same happens for $m_{01} > 2 \, R_y^{-1}$.
  Several caveats exist, though. For one, the bounds depend very
  sensitively on the choice for the galactic DM profile; for DM
  particles as heavy as we consider and the somewhat nonstandard
  couplings that it has, the uncertainties in the profile are quite
  significant. Moreover, choosing a moderately large value of $\Lambda
  R_y$ further suppresses such cross sections. It should be noted here
  that the embedding in the warped dimension essentially removes the
  stringent constraints on $\Lambda R_y$ that were operative in the
  canonical UED scenario.  Nonetheless, this would remain a
  particularly interesting channel and is the most likely theater for
  a critical examination of the paradigm. Naively, it might seem that
  the very same coupling (of the DM to the photon) would lead to
  severe modification in CMB observations\cite{Wilkinson:2013kia} and
  long-distance observations, such as those of blazars and
  quasars. However, once again, the scattering of such photons by the
  intervening DM would need to proceed through $t$-channel graviton
  exchanges, and given the size of the couplings, are of little
  concern.

  Finally, we comment on the prospects at the LHC. While the
  presence of the gravitons (particularly, the $h_{01}$), has allowed
  us the upper limit of $R^{-1} \sim 1400 \gev$ (operative in the
  canonical minimal UED) from considerations of the relic density, it
  is not that one can allow of arbitrarily large $R_y^{-1}$. As
  Fig.\ref{fig:param} suggests, larger values of $R_y^{-1}$ typically
  require larger $\epsilon$ values, thereby calling into question the
  semiclassical treatment. Indeed, such considerations restrict one to
  $R_y^{-1} \lapp 3.5 \tev$ thereby bringing the model wholly into the
  reach of the LHC.

\acknowledgments

DC acknowledges partial support from the European Union's Horizon 2020
research and innovation program under Marie Sk{\l}odowska-Curie grant
No 690575. DS would like to thank UGC-CSIR, India for financial assistance.

\bibliographystyle{JHEP}
\bibliography{reference}
\end{document}